\documentclass[twocolumn,titlepage]{revtex4}
\usepackage[english]{babel}
\usepackage{setspace} 
\usepackage{amsmath}
\usepackage{multirow}
\usepackage{graphicx}

\begin{document}
\title{Brownian colloidal particles: Ito, Stratonovich or a different stochastic interpretation}

\author{J. M. Sancho}
\affiliation{Universitat de Barcelona, Departament d'Estructura i Constituents de la Mat\`eria,\\
Mart\'i i Franqu\'es, 1. 08028 Barcelona. Spain}

\begin{abstract}
Recent experiments on Brownian colloidal particles have been studied theoretically in terms of overdamped Langevin equations with multiplicative white noise using an unconventional stochastic interpretation. Complementary numerical simulations of the same system  are well described using the conventional Stratonovich interpretation. Here we address this dichotomy from a more generic starting point: the underdamped Langevin equation and its corresponding Fokker--Planck equation.

\end{abstract}


\maketitle


{\bf Introduction}.

In a recent controversy it has been claimed, based on a well controlled experiment,
that  colloidal particles  obey an overdamped  Langevin equation with multiplicative
noise that
can not be interpreted using the standard Ito or the standard Stratonovich interpretations, 
but that instead requires an altogether different interpretation \cite{Volpe}.
This particular  choice has in turn
been criticized and, based on numerical simulations, the claim has been made that
the proposed overdamped Langevin equation
must instead be corrected by including a new drift term, and that this
augmented equation requires
the Stratonovich interpretation \cite{Mannella}.
Here we will show that both positions are correct, but that the latter procedure
can be associated with a systematic analytical derivation from the full Langevin equation.

In the experiment \cite{Volpe} electrically charged dielectric colloidal spheres move in
a solvent. The interaction forces have been demonstrated to be
described by a conservative potential $V(z)$ in the vertical
direction $z$ which includes a repulsive potential from  a surface at $z=0$ and the
appropriate gravitational contributions.  The particles are also subject to 
thermal fluctuations and a position-dependent friction $\lambda(z)$.  Both
$V(z)$ and $\lambda(z)$ have been checked and are well controlled. The particle's trajectory
is analyzed using different refined techniques at $nm$ and $ms$ scales.

As this is an experimentally well prepared and theoretically well described system
in statistical physics,
we can unequivocally state the main facts: the particle motion is driven by
the classical dynamical equation of a damped Brownian particle and its steady
state is governed by
the Boltzmann distribution  corresponding to the system Hamiltonian. Thus the
dynamical equation is,

\begin{equation}
 {\ddot z}= - \lambda(z) {\dot z} - V'(z) + g(z) \xi(t),
\label{langevin1}
\end{equation}
where $\xi(t)$ is a Gaussian white noise of zero mean and correlation function
\begin{equation}
 \langle \xi(t) \xi (t') \rangle = 2\delta (t-t'),
\label{corr}
\end{equation}
and for simplicity we have set the particle mass to unity.

The multiplicative function $g(z)$ and the friction $\lambda(z)$ obey the
fluctuation dissipation relation,
\begin{equation}
  g(z)^2 = k_B T \lambda(z)=\lambda^2(z) D_{\bot}(z),
\label{FD}
 \end{equation}
where the notation in \cite{Volpe} has been used in the last term to stress the
relation  with the diffusion coefficient $D_{\bot}(z)$.

One can check that the stationary distribution corresponding to the Fokker--Planck
equation for the probability density $P_{st}({\dot z}, z)$ is the Boltzmann distribution,
\begin{equation}
P_{st}({\dot z}, z) \sim  \exp \left(- \frac{ {\dot z}^2/ 2 + V(z)}{k_B T} \right).
\label{B}
\end{equation} 
This is the correct starting point to analyze  the experimental observations and to clarify the
theoretical stochastic description. 

When the friction dominates over the inertial forces one can approximate the second order differential equation (\ref{langevin1}) by one of first order (overdamped Langevin equation)
where the acceleration term does not appear.
However, a first order Langevin equation with multiplicative white noise requires a
stochastic interpretation
\cite{Stratonovich} characterized by the value of the parameter $\alpha$ (the role of this parameter $\alpha$ is detailed in the appendix).
The controversy occurs because of the claim in \cite{Volpe} that
this physical system can be described by overdamped Langevin dynamics with
multiplicative noise that does not adhere to the well known stochastic interpretations, that
is,
to either the Ito ($\alpha=0$) or the Stratonovich  ($\alpha=1/2$) interpretation.  Instead, they claim that 
an unconventional  stochastic interpretation with $\alpha=1$ is required.
On the other hand, it is claimed in \cite{Mannella} on the basis of numerical
simulations of the same system that the correct interpretation instead requires a
modification of the overdamped Langevin equation, and that the resulting equation must be
interpreted according to Stratonovich, as commonly accepted in the physics community.

In this brief report it will be proved that both positions represent the same physical
content, but that the Stratonovich proposal of  \cite{Mannella}  can in fact
be derived from a careful adiabatic elimination of the velocity in the
full Langevin equation that includes the inertial contribution. The explicit derivation
of this result makes use of calculations concerning stochastic calculus that were carried out
three decades ago \cite{Sancho}. 


We next present a unified analysis of both approaches. Technical details are explained in the appendix.\\

{\bf The stochastic analysis}.

First it is worth to clarify that there is neither any doubt nor a controversy  about the interpretation of the infradamped stochastic differential
equation (\ref{langevin1}). We can operate it using the standard rules of calculus.

The controversy starts when one wants to write the overdamped Langevin equation corresponding to (\ref{langevin1})-(\ref{FD}). In this regard, it must be stressed that the overdamped limit is {\em not} obtained by simply setting
the acceleration equal to zero,
\begin{equation}
 {\dot z} = -\frac{V'(z)}{\lambda(z)}  + \frac{g(z)} {\lambda(z)} \xi(t),
\label{langevin2}
\end{equation}
which is precisely the starting point in the analysis of Ref.  \cite{Volpe}.

For a better comparison of the two proposals \cite{Volpe,Mannella} they can be cast in the single overdamped equation,

\begin{equation}
{\dot z} = -\frac{V'(z)}{\lambda(z)} - \frac{\beta k_B T}{2} \frac{\lambda'(z)}{\lambda^2(z)} + \sqrt{\frac{k_B T} {\lambda(z)}} \xi(t),
\label{langevin3}
\end{equation}
where the correlation function of the noise is given by (\ref{corr}).
Here $\beta$ is an auxiliary parameter to facilitate comparison of the different Langevin
equations under consideration.  Use of (\ref{FD}) has been made to simplify the notation.
The Langevin equation proposed in \cite{Volpe} corresponds to the choice  $\beta_{V}=0$,
while that of \cite{Mannella} corresponds to the choice  $\beta_{MM} =1$. This is not enough and the stochastic interpretation of (\ref{langevin2}) or (\ref{langevin3}) matters here.

Nevertheless the Fokker--Planck equation corresponding to a Langevin equation is a much better reference frame to discuss these matters. This equation has the following standard form,
\begin{equation}
 \frac{\partial P}{\partial t } = - \frac{\partial}{\partial z} K_1(z) P + \frac{1}{2} \frac{\partial^2}{\partial z^2} K_2(z) P,
\label{FP1}
\end{equation}
where $K_1(z), K_2(z)$ are the first two differential moments, which can be calculated
from the explicit algorithm (\ref{algo2}) with the results (see the appendix for details),
\begin{eqnarray}
 K_1(z)&=& -\frac{V'(z)}{\lambda(z)}- \frac{ k_B T}{2} (\beta + 2\alpha) \frac{\lambda'(z)}{\lambda^2(z)} ; 
\nonumber\\
 K_2(z) &=& 2\frac{k_B T}{ \lambda(z)}.
\label{ks}
\end{eqnarray}

Then the associated Fokker--Planck equation corresponding to (\ref{langevin3}) can be written in the form,
\begin{eqnarray}
\frac{\partial P}{\partial t}=  \frac{\partial}{\partial z} \frac{1}{\lambda(z)} \left[ V'(z) +
k_B T \frac{\partial }{\partial z} \right]P + \nonumber
\\ 
+\frac{\partial}{\partial z} \frac{k_B T}{2} \frac{\lambda'(z)}{\lambda^2(z)} \left( \beta + 2\alpha - 2 \right) P.  
\label{FP4}
\end{eqnarray}

Now it is time to discuss both interpretations. In Ref. \cite{Volpe} it is claimed that the parameter $\alpha_{V}=1$ with $\beta=0$ is the correct choice,
while in Ref. \cite{Mannella} the choice is the standard Stratonovich one, $\alpha_{MM} = 1/2$ but with $\beta=1$.

Now it is quite clear that the two options discussed above cause the 
last term term to vanish because all of them fulfill the condition
\begin{equation}
 \beta + 2 \alpha = 2,
\label{betaalpha}
\end{equation}
and accordingly they obey the same
Fokker--Planck equation,
\begin{equation}
 \frac{\partial P}{\partial t}=  \frac{\partial}{\partial z} \frac{1}{\lambda(z)} \left[ V'(z) +
k_B T \frac{\partial }{\partial z} \right]P.
\label{FP3}
\end{equation}
Furthermore, this means that they  reflect the 
same physics contained in the now reduced Boltzmann distribution, 
\begin{equation}
P_{st}(z) \sim e^{- V(z)/k_BT}, 
\end{equation}
which is a solution of (\ref{FP3}). 

We can conclude that all the possible models and stochastic interpretations satisfying (\ref{betaalpha}) will give the same physics. Moreover  
for those readers attached to the Ito interpretation, the choice would be $\beta=2$ and $\alpha=0$, and
the readers interested in having their own choice they can select a pair of $\beta$ and $\alpha$ fulfilling (\ref{betaalpha}) .

This result closes the immediate controversy, but a question remains open:
Does a correct choice exist?  The answer is yes, and it is based on the adiabatic
elimination of the velocity from the full Langevin equation description (\ref{langevin1})-(\ref{corr}).  

The adiabatic elimination of the velocity variable in the full inertial Langevin
equation was carried out following rigorous mathematical procedures in \cite{Sancho}.
The key point is an expansion in the $\lambda(z)$ as a large quantity, which leads to
the overdamped approximation to the problem. This is manifest in the final
form of the Fokker--Planck equation, which is of order $\lambda(z)^{-1}$ in the drift
and diffusion terms. The final result is unequivocally that the physically correct overdamped
equation is (\ref{langevin3}) which corresponds to the proposal of Ref. \cite{Mannella}, that is,
$\beta=1$ along with the choice $\alpha= 1/2$ corresponding with the Stratonovich interpretation.
The technical details are indeed cumbersome but available to the interested reader,
in \cite{Sancho} or upon request to the author.  The way in which the final overdamped
equation depends on the physical constrains during the adiabatic elimination procedure is
also studied in \cite{Kupferman}. A thermodynamic view of the different stochastic interpretations is discussed in \cite{Sokolov}. \\

{\bf Force and drift.} 

Another discussion in \cite{Volpe, Mannella} is devoted to
the correct evaluation of the force felt by the
colloidal particles. If we know all the terms in the starting Langevin equation,
it is obvious that the only physical force of the problem comes from the potential
energy $f(z)=- V'(z)$, which can be extracted experimentally from the steady state
Boltzmann distribution. Another question is the evaluation of the drift on the particle position $z$ (which
should not be confused  with  the mechanical force) in the overdamped description.
According to most of the literature, the drift is just the quantity $K_1(z)$ which
appears in the Fokker--Planck description. In our case this is given explicitly by
\begin{equation}
 K_1(z)= -\frac{V'(z)}{\lambda(z)} - k_B T \frac{\lambda'(z)}{\lambda^2(z)} = \frac{f(z)}{\lambda(z)} + \frac{d}{dz} D_{\bot}(z),
\end{equation}
which depends neither on $\beta$ nor on the stochastic interpretation parameter $\alpha$ because of (\ref{betaalpha}).
The drift can be obtained from experimental data on $z(t)$ trajectories using its definition (see
(\ref{K1}) in the appendix). 
 
We thus conclude that matters have been clarified without the need to introduce
new stochastic interpretations to understand this well controlled physical system
either experimentally \cite{Volpe} or numerically \cite{Mannella}. The keystone is the connection between the infradamped equation (\ref{langevin1}) and the corresponding overdamped equation (\ref{langevin3}), and not the more intuitive overdamped equation (\ref{langevin2}).

We acknowledge financial support from projects  FIS2009-13360-C03-01
(Ministerio de Educaci\'on y Ciencia of Spain). Fruitful discussions with 
Profs. K. Lindenberg and R. Mannella are also acknowledged. 

{\bf Appendix.}  Here we present the derivation of the Fokker--Planck equation
for the generic Langevin equation with multiplicative noise \cite{Sagues},
\begin{equation}
 {\dot z} = A(z) + B(z) \xi(t),
\label{langevingen}
\end{equation}
where the noise  correlation is given by (\ref{corr}). We start by deriving a 
first order algorithm for a time step $\Delta t$.  Formal integration of (\ref{langevingen}) leads
to
\begin{equation}
z(t+\Delta t) = z(t) + \int_t^{t+\Delta t}  A(z(t')) dt'+  \int_t^{t+\Delta t} B(z(t')) \xi(t') dt'.
\label{algo1}
\end{equation}
An expansion of the first integral gives  $A(z(t)) \Delta t$, but the second integral is
not unequivocally defined
\begin{eqnarray}
\int_t^{t+\Delta t} B(z(t')) \xi(t') dt' \equiv \nonumber
\\
 B\left[ (1-\alpha) z(t) + \alpha z(t+\Delta t) \right] \Delta W(t).
\label{interpretation}
\end{eqnarray}
because it does depend on a parameter $\alpha\,\epsilon \, [0,1]  $ which specifies the stochastic interpretation of the Langevin equation.
$\Delta W(t)$
is the Wiener increment,
\begin{equation}
 \Delta W(t) = \int_t^{t+\Delta t} \xi(t') dt',
\end{equation}
which is a Gaussian precess with zero mean value and second moment
\begin{equation}
 \langle \Delta W(t)^2 \rangle = 2 \Delta t.
\end{equation}

The next step is to write an algorithm for (\ref{algo1}) to first order in $\Delta t$, 
which implies an expansion of the stochastic integral. Using the fact that to lowest
order $z(t + \Delta t)$ in (\ref{interpretation}) can be approximated by
\begin{equation}
 z(t + \Delta t) = z(t) + B(z(t)) \Delta W(t) + \ldots,
\end{equation}
we get the explicit form of the algorithm for any stochastic interpretation $\alpha$,
\begin{eqnarray}
  z(t + \Delta t) &=& z(t) + A(z(t)) \Delta t + B(z(t)) \Delta W(t) +
\nonumber\\
 &&\alpha B(z(t)) B'((z(t)) \Delta W(t)^2 + \ldots.              
\label{algo2}
\end{eqnarray}

The Fokker--Planck equation (\ref{FP1}) for $P(z,t)$ can be obtained
from the Kramers--Moyal expansion \cite{Stratonovich}, where the quantities $K_1(z)$ and $K_2(z)$ 
are obtained as follows.
 Defining the increment $\Delta z = z(t + \Delta t) - z(t) $,
we find
\begin{eqnarray}
 K_1(z) &=& \lim_{\Delta t \rightarrow 0} \frac{ \langle \Delta z \rangle}{\Delta t} = A(z) + 
2 \alpha B(z) B'(z),
\nonumber\\
 K_2(z) &=& \lim_{\Delta t \rightarrow 0} \frac{ \langle (\Delta z )\rangle^2 }{\Delta t} = 2 B(z)^2.
\label{K1}
\end{eqnarray}
These are the two quantities which appear in the Fokker--Planck equation (\ref{FP1}). 
Using now the equation (\ref{langevin3}) we get the explicit expressions for $A(z)$ and $B(z)$,
\begin{equation}
 A(z) = -\frac{V'(z)}{\lambda(z)} - \frac{\beta k_B T}{2} \frac{\lambda'(z)}{\lambda^2(z)};  
B(z) = \sqrt{\frac{k_BT}{\lambda(z)}},
\end{equation}
which are substituted in the previous equations to obtain (\ref{ks}).

\end{document}